\documentclass[prx,twocolumn,floatfix,superscriptaddress,longbibliography]{revtex4-1}

\usepackage[T1]{fontenc}
\usepackage{url}
\usepackage{ifthen}
\usepackage[cmex10]{amsmath} 
\usepackage[utf8]{inputenc} 
\usepackage{amsmath} 
\usepackage{graphicx} 
\usepackage[usenames]{color} 
\usepackage{amssymb} 
\usepackage{supertabular}
\usepackage{subfigure}
\usepackage{physics}
\usepackage[export]{adjustbox}
\usepackage{mathtools}
\usepackage[ruled,vlined]{algorithm2e}
\usepackage{float}
\usepackage{tabularx}
\usepackage{array, makecell}
\usepackage[colorlinks=true,citecolor=blue,linkcolor=red,urlcolor=black]{hyperref}

\newtheorem{definition}{Definition}
\newtheorem{example}{Example}

\newcommand{\FF}{\mathbb{F}}

\newcommand{\ZZ}{\mathbb{Z}}
\newcommand{\fanf}{f^{(\mathbb{F})}}
\newcommand{\fnnf}{f^{(\mathbb{Z})}}

\newcommand{\mq}{\mathsf{MQ}}

\begin{document}

\title{Solving systems of Boolean multivariate equations with quantum annealing}

\author{Sergi Ramos-Calderer}
\affiliation{Quantum Research Centre, Technology Innovation Institute, Abu Dhabi, UAE.}
\affiliation{Departament de F\'isica Qu\`antica i Astrof\'isica and Institut de Ci\`encies del Cosmos (ICCUB), Universitat de Barcelona, Mart\'i i Franqu\`es 1, 08028 Barcelona, Spain.}

\author{Carlos Bravo-Prieto}
\affiliation{Quantum Research Centre, Technology Innovation Institute, Abu Dhabi, UAE.}
\affiliation{Departament de F\'isica Qu\`antica i Astrof\'isica and Institut de Ci\`encies del Cosmos (ICCUB), Universitat de Barcelona, Mart\'i i Franqu\`es 1, 08028 Barcelona, Spain.}

\author{Ruge Lin}
\affiliation{Quantum Research Centre, Technology Innovation Institute, Abu Dhabi, UAE.}
\affiliation{Departament de F\'isica Qu\`antica i Astrof\'isica and Institut de Ci\`encies del Cosmos (ICCUB), Universitat de Barcelona, Mart\'i i Franqu\`es 1, 08028 Barcelona, Spain.}

\author{\\ Emanuele Bellini}
\affiliation{Cryptography Research Centre, Technology Innovation Institute, Abu Dhabi, UAE.}

\author{Marc Manzano}
\affiliation{Sandbox@Alphabet, Mountain View, CA, USA.}
\affiliation{Electronics and Computing Department, Faculty of Engineering, Mondragon Unibertsitatea, Mondragon, Spain.}

\author{Najwa Aaraj}
\affiliation{Cryptography Research Centre, Technology Innovation Institute, Abu Dhabi, UAE.}

\author{Jos\'e I. Latorre}
\affiliation{Quantum Research Centre, Technology Innovation Institute, Abu Dhabi, UAE.}
\affiliation{Departament de F\'isica Qu\`antica i Astrof\'isica and Institut de Ci\`encies del Cosmos (ICCUB), Universitat de Barcelona, Mart\'i i Franqu\`es 1, 08028 Barcelona, Spain.}
\affiliation{Centre for Quantum Technologies, National University of Singapore, Singapore.}

\begin{abstract}
Polynomial systems over the binary field have important applications, especially in symmetric and asymmetric cryptanalysis, multivariate-based post-quantum cryptography, coding theory, and computer algebra. In this work, we study the quantum annealing model for solving Boolean systems of multivariate equations of degree 2, usually referred to as the Multivariate Quadratic problem. We present different methodologies to embed the problem into a Hamiltonian that can be solved by available quantum annealing platforms. In particular, we provide three embedding options, and we highlight their differences in terms of quantum resources. Moreover, we design a machine-agnostic algorithm that adopts an iterative approach to better solve the problem Hamiltonian by repeatedly reducing the search space. Finally, we use D-Wave devices to successfully implement our methodologies on several instances of the Multivariate Quadratic problem.
\end{abstract}

\maketitle

\section{Introduction}


Adiabatic quantum computation is a universal quantum computation scheme \cite{farhi2000quantum} where a quantum system is prepared in the ground state of an easy to prepare Hamiltonian and evolved towards a Hamiltonian that encodes the solution of a problem in its ground state. If the evolution is performed adiabatically, the quantum system still remains in its instantaneous ground state and the problem will be solved. Quantum annealers are special-purpose devices based upon the principles of adiabatic quantum computation that work with simpler Hamiltonains and more relaxed evolution times. However, these devices are believed to provide an edge when solving classical satisfiability problems by leveraging quantum phenomena and are easier to control and scale up for larger, real-life problems~\cite{finnila1994quantum, brooke1999quantum, johnson2011quantum, perdomo2012finding, mandra2016strengths, benedetti2017quantum, khoshaman2018quantum, benedetti2018quantum, ding2019towards, perdomo2019readiness, wilson2021quantum}. Interestingly, quantum annealers on the order of thousands of qubits are already commercially available by D-Wave~\cite{dwave}. 


The quantum annealing approach to quantum computing is a research topic that can be relevant on many research problems in a variety of scientific fields. Motivated by this idea, we investigate the possibility of using D-Wave quantum annealer for a fundamental problem in computer science: solving systems of multivariate polynomial equations over the binary field.
If all polynomials in the system are linear, 
then the system can be efficiently solved, for instance, by Gaussian elimination.
The problem is easy also when the system is 
either underdetermined (much fewer equations than variables), overdetermined (much more equations than variables), or sparse
(the number of terms is linear with respect to the number of variables). 
However, the problem is known to be NP-hard already for generic quadratic systems
\cite{lewis1983computers}. 
Moreover, assuming the exponential time hypothesis
\cite{impagliazzo2001complexity}, 
there exists no sub-exponential time (worst-case) algorithm for this problem.

Polynomial systems over the binary field can be used to perform algebraic cryptanalysis 
\cite{bard2009algebraic} potentially against any cipher.
Moreover, in the case of degree 2 polynomials, the problem, usually referred to as the Multivariate Quadratic ($\mq$) problem, 
has important applications in post-quantum cryptography, 
since several post-quantum schemes
exist basing their security on its difficulty to be solved \cite{rainbowround3nist, gemmsround3nist}. 
For these reasons, there is a spreading interest in the scientific community to find new algorithms to solve the $\mq$ problem, 
both in the classical and quantum computation model.
The former case has been extensively studied,
see for example \cite{Bard2007,mou:tel-01110887,ullah2012,EDER2017719} for comprehensive surveys on the most effective algorithms.
Regarding the latter,
a detailed analysis of the required qubits and time for a Grover's algorithm approach is presented in Ref.~\cite{schwabe2016solving}. Building upon the previous work, 
the authors of Ref.~\cite{pring2018exploiting} 
demonstrate that by applying preprocessing 
the computational load on the quantum computer can be reduced and, 
in a generalization of the multi-target search for single targets, 
the efficiency of the basic quantum search oracle for the $\mq$ problem over the binary field can be improved. In Ref.~\cite{faugere2017fast}, the authors present a quantum version of BooleanSolve \cite{bardet2013complexity}, which is currently the fastest asymptotic algorithm for classically solving systems of non-linear Boolean equations, that takes advantage of Grover’s quantum algorithm. 
Note that Ref.~\cite{bernstein2018asymptotically} also proposed a new Gr\"obner-based quantum algorithm for solving quadratic equations with a complexity comparable to QuantumBooleanSolve (we refer to Ref.~\cite{faugere2017fast} for further details).
Finally, in Ref.~\cite{biasse2021framework}, 
the authors show how to reduce the use of quantum RAM and circuit complexity by delegating some precomputations to a classical computer.

Note that all the above quantum techniques have only considered the use of universal fault-tolerant quantum computers. Therefore, the aforementioned methods can not be implemented on current quantum computers without error correction. In contrast, the use of quantum annealers to solve systems of multivariate equations over a finite field is still unexplored, 
and in this work, we try to fill this gap. 

In this paper, we explore how quantum annealing 
can be used 
for 
solving multivariate systems of quadratic equations over binary fields, namely, the $\mq$ problem.
We present different methodologies to translate the $\mq$ problem into a Hamiltonian that can be solved by a quantum annealer. 
To support our proposal, we provide results obtained by running examples on D-Wave's Advantage quantum device \cite{dwave}, a commercially-available quantum computer. Our approach takes into consideration the decomposition of multi-qubit terms up to at most two-qubit interactions, which is a constraint of the underlying architecture of near-term quantum annealing devices.

Our work highlights the main obstacle of mapping Boolean systems of equations into Hamiltonian ground states, which is the fact that the inherent correlations in operations over the binary field have to be encoded into a real Hamiltonian, at the cost of an overhead. We present two alternative methods to circumvent the exponential overhead in quantum memory that the naive transformation would incur.
Moreover, we also introduce a new iterative approach that aids quantum annealing devices in finding the ground state of the Hamiltonian by repeatedly shrinking the search space using information gained in previous executions of the system. This method allowed us to successfully solve small instances of the $\mq$ problem using current D-Wave devices.

This paper is organized as follows. In Sec. \ref{sec:preliminaries}, we introduce the background required for this work. Next, in Sec. \ref{sec:formalizing} we present our methodologies to translate a given $\mq$ problem into a Hamiltonian in the context of quantum annealing and analyze the needed resources.  Finally, in Sec. \ref{sec:experiments} and Sec. \ref{sec:conclusions} we present the results of our experiments with D-Wave and the conclusions of this work, respectively.






\section{Preliminaries}
\label{sec:preliminaries}

We denote by $\mathbb{F}_q$ the finite field with $q$ elements. $\mathbb{F}_q^n$ is the set of all vectors of length $n$, viewed as an $\mathbb{F}_q$-vector space. 
For compactness, we sometimes denote with 
$\Vec{x}$ the vector $(x_1, \dots, x_n)$.

The $\mq$ problem is defined as follows. The input of the problem consists of $m$ quadratic polynomials $p_1(x_1, \dots, x_n), \dots, p_m(x_1, \dots, x_n) \in \mathbb{F}_q[x_1, \dots, x_n]$ in $n$ variables $x_1, \dots, x_n$ and coefficients in a finite field $\mathbb{F}_q$.
The output of the problem is given by the set of $(a_1, \dots, a_n) \in \mathbb{F}_q^n$ for which 
$p_i(a_1, \dots, a_n) = 0$ for all $i = 1, \ldots, m$.
The vector $(a_1, \dots, a_n)$ is called a \emph{solution} of the \emph{system of equations}
\begin{equation}\label{equ:system}
p_i(x_1, \dots, x_n) = 0, \quad i = 1,\dots,m .
\end{equation}
Three variants of the $\mq$ problem can be defined. (1) The \emph{Decision} variant asks to determine if Eq.~\eqref{equ:system} has a solution.
(2) The \emph{Search} variant asks to find a solution of Eq.~\eqref{equ:system}, if there is one. 
(3) The \emph{Exhaust} variant asks to find all solutions of Eq.~\eqref{equ:system}.

The decision variant of $\mq$ is known to be an NP-complete problem
\cite{fraenkel1979complexity}.
It is easy to observe that solving the search variant also solves the decision one.
On the other hand, by iteratively guessing each variable, it is possible to solve the search variant by solving the decision variant at most $n$ times.

For practical purposes, one is usually interested in the search variant, and sometimes in the exhaust variant.
From now on, 
unless it is stated otherwise, we write $\mq$ problem to refer to its search variant. 
Furthermore, we will focus on the Boolean case, i.e., where $q=2$.
In this case, polynomials are called Boolean polynomials, and 
the corresponding unique map $f$ from $\mathbb{F}_2^n$ to $\mathbb{F}_2$ is called 
a Boolean function.
It is common to refer to the Boolean polynomial as the Algebraic Normal Form (ANF) of $f$, 
which we indicate with $\fanf$.
In this work, we will also need another representation of $f$
called the Numerical Normal Form (NNF), which we indicate with $\fnnf$.
\begin{definition}\label{defNNF}
 Let $f$ be a Boolean function on $\FF_2^n$ taking values in the integer ring $\ZZ$. 
 We call the \emph{Numerical Normal Form (NNF)} of $f$ 
 the following expression of $f$ as a polynomial:
 $$
 f(x_1,\ldots,x_n) 
 = 
 \sum_{u \in \FF_2^n}\lambda_u (\prod_{i=1}^{n}x_i^{u_i}) 
 = 
 \sum_{u \in \FF_2^n}\lambda_{u}X^u\,,
 $$
 with $\lambda_{u} \in \ZZ$ and $u=(u_1,\ldots,u_n)$.
\end{definition}
With abuse of notation and when clear from the context, 
we sometimes write $f = \fanf = \fnnf$, 
and we indicate with $+$ both the addition over $\FF_2$ or over another field or ring. 
Given a Boolean function, its ANF and NNF are unique.
It is also worth noting that, in general, if a Boolean function in ANF has about $k$ terms (i.e., nonzero coefficients), then its corresponding NNF will contain about $2^k$ terms (see Ref. \cite{bellini2018deterministic} for detailed proof). 
As described in this work, this significant increase of terms turns out to be the main obstacle when trying to solve Boolean polynomial systems using annealing evolution.
We finally refer to \cite{carlet10boolean} for an exhaustive introduction to Boolean functions.

\begin{example}
An example of a quadratic Boolean polynomial system with
$n = 4$ variables and 
$m = 4$ equations
is given below:
{\footnotesize
\begin{align*}
    \begin{cases}
    x_1 x_2 + x_1 x_3 + x_1 x_4 + x_1 + x_2 x_3 + x_2 x_4 + x_2 + x_3 x_4 + x_4 = 0 \\
    x_1 x_2 + x_1 x_3 + x_2 + x_3 x_4 + x_3 = 0 \\
    x_1 x_2 + x_1 x_3 + x_2 x_3 + x_2 + x_3 + x_4 = 0\\ 
    x_1 x_3 + x_2 x_4 + x_4 + 1 = 0 \\
    \end{cases}
\end{align*}
}
The polynomials are given in Algebraic Normal Form and the only solutions of the system are the two binary vectors 
$(1, 0, 1, 0)$ and $(0, 0, 1, 1)$.
For example, the Numerical Normal Form of 
$\fanf = x_1 x_3 + x_2 x_4 + x_4 + 1$ (note that the addition is over $\FF_2$) 
is given by 
$\fnnf = -2 x_1 x_2 x_3 x_4 + 2 x_1 x_3 x_4 - x_1 x_3 + x_2 x_4 - x_4 + 1$
(note that the addition is over the integer ring $\ZZ$).
\end{example}

\section{Formalizing the problem in a quantum annealer}
\label{sec:formalizing}

In the early 2000s, Farhi et al. proposed a new universal quantum computation model based on the quantum adiabatic theorem \cite{farhi2000quantum, farhi2001quantum}. The so-called adiabatic quantum computation model was shown to be polynomially equivalent to the quantum gate-based model proposed by Deutsch in 1989 \cite{deutsch1989, aharonov2008}, and is one of the most promising models of quantum computing due to its natural robustness against errors~\cite{childs2001robustness}. The adiabatic theorem guarantees that if the Hamiltonian that dictates the energy of a quantum system is modified slowly enough, a quantum state will remain in its instantaneous ground state during the evolution~\cite{born1928, kato1950adiabatic}. 
This implies that we can encode the solution of a hard problem, the $\mq$ problem in this case, into the ground state of a problem Hamiltonian $H_p$, and then, starting from an easy-to-prepare ground state of an initial Hamiltonian $H_0$, drive the system slowly to the problem Hamiltonian to then measure its solution.

A less restrictive, more hardware-friendly, technique to solve classical problems is quantum annealing. 
Quantum annealing also follows the evolution of a quantum Hamiltonian in order to find low-energy configurations of the system but does not demand adiabatic evolution and forgoes universality. Devices such as D-Wave are quantum annealers, and while not being universal quantum computers due to their limitations, are still useful for solving hard optimization problems \cite{kadowaki1998quantum, das2008}. Hard classical problems that can be codified to a problem Hamiltonian by only using the computational basis of the system, as is the case for the $\mq$ problem tackled in this paper, are ideal for these available quantum annealers.

We want to solve the $\mq$ problem defined in Sec. \ref{sec:preliminaries}, where we are given a set of $m$ quadratic polynomials $p_1(x_1,\ldots x_n),\ldots p_m(x_1, \ldots x_n)$ over the binary field and we are tasked with finding $\Vec{x}$ so that all $\Vec{p}$ are equal to zero, via quantum annealing. Therefore, we need to create a Hamiltonian with a ground state that encodes the solution to this problem.

\subsection{Direct embedding}
\label{sec:direct}


A first direct approach is penalizing with positive energy each of the equations $p_i(\Vec{x})$ that is not fulfilled. The corresponding problem Hamiltonian can be constructed as 
\begin{equation}
    H_p = \sum_{i=1}^mp_i(\Vec{x})\,,
\label{eq:Hp}
\end{equation}
as it contributes with positive energy if the input bits for $p_i(\Vec{x})$ do not result in a zero solution.
Usually, the polynomials $p_i(\Vec{x})$ in Eq. \ref{eq:Hp} are given in ANF since bitwise operations are performed over the binary field $\FF_2$. However, the quantum Hamiltonian we can encode into a quantum annealer device does not function with binary algebra, each positive term only adds more energy to the final state. Therefore, each polynomial $p_i(\Vec{x})$ has to be given in its NNF. This transformation can be obtained by recursively applying the change
\begin{equation}
    (x_i + x_j) \longrightarrow x_i + x_j - 2 x_i\cdot x_j
\label{eq:ANFNNF}
\end{equation}
to the original ANF equations, where, with abuse of notation, the symbol $+$ on the left is the addition over $\FF_2$, 
while the symbols $+$ and $-$ on the right are the regular addition and subtraction over the integer ring. 
This transformation introduces multi-qubit interaction terms (i.e., terms of degree greater than 1) that were not present in the ANF of $p_i(\Vec{x})$. In general, all combination of monomials present in the ANF of $p_i(\Vec{x})$ will appear in the NNF. Keep in mind that, in a binary field, it holds that $x^2=x$ (since the values of the variables $x_i$ are either $0$ or $1$), there are no powers in the monomials of the ANF or the NNF.
This transformation gives rise to a different issue, the chip architecture of currently available quantum annealers only allows for two-qubit interactions. Thus, to run a quantum annealing protocol on a real device, the interactions of the Hamiltonian have to be reduced. For a general many-body Hamiltonian, its interactions can be reduced to two-body using perturbation theory by adding ancilla qubits \cite{bravyi2008quantum, jordan2008perturbative, cao2015hamiltonian}. If all the problem Hamiltonian parts share the same basis, as is the case for a classical Hamiltonian such as ours, the reduction can be performed without perturbation theory \cite{biamonte2008non, babbush2013resource}. This reduction method yields a new Hamiltonian with a different energy spectrum but equal ground state and energy, therefore not altering the solution of the problem and is the one we follow for the direct embedding.
The method consists of exchanging a two-qubit interaction by an ancilla, reducing by one the order of the interaction. A penalty function is then introduced to the Hamiltonian that adds energy when the value of the ancilla is not equal to the product of the original two qubits. The penalty function can be written as 
\begin{equation}
    s(x_i, x_j, x_{ij}) = 3x_{ij}+x_ix_j-2x_ix_{ij}-2x_jx_{ij}\,,
\end{equation}
where $x_{ij}$ is the label given to the ancillary qubit that is substituted. It can be seen that $s(x_i, x_j, x_{ij})=0$ if $x_ix_j=x_{ij}$ and $s(x_i, x_j, x_{ij})\geq 1$ otherwise. This keeps the ground state and energy unchanged.

Furthermore, a single ancilla $x_{ij}$ can be used for all terms in the Hamiltonian, where the term $x_ix_j$ appears.
This is achieved by applying the substitution 
\begin{align}\label{eq:substitution}
    \sum_K\alpha_{ijK}&x_ix_jx_K\longrightarrow\\\nonumber
    &\sum_K\left(\alpha_{ijK}x_{ij}x_K+\left(1+\abs{\alpha_{ijK}}\right)s(x_i, x_j, x_{ij})\right)\,,
\end{align}
where the index $K$ is the product of multiple other variables in all terms where $x_ix_j$ is present. It can be shown that this transformation also yields a Hamiltonian with the same ground state \cite{babbush2013resource}.
When this procedure is used to reduce large multi-qubit terms, the resulting final Hamiltonian will have large coefficients. This introduces a problem for real-life implementation since the machine precision for coefficients of quantum annealing devices such as D-Wave's is limited. The quantum annealer developed by D-Wave scales the given coefficients between $[-1, 1]$ when introducing them to the machine, so small coefficients can vanish when translated into weights in the presence of other large parameters.
An alternative transformation is proposed in Ref. \cite{babbush2013resource} with the aim of reducing the precision needed for the control of the device. Introducing the term
\begin{equation}
    \delta_{ij}=\max\left(\sum_{K, \alpha_{ijK}>0}\alpha_{ijK}, \sum_{K, \alpha_{ijK}<0}-\alpha_{ijK}\right)\,,
\end{equation}
the substitution given in Eq. \eqref{eq:substitution} can be rewritten as
\begin{equation}
    \begin{split}
    \sum_K\alpha_{ijK}x_ix_jx_K &\longrightarrow  \\ &\sum_K\alpha_{ijK}x_{ij}x_K+\left(1+\delta_{ij}\right)s(x_i, x_j, x_{ij})\,,
    \end{split}
\label{eq:control}
\end{equation}
while still keeping the desired ground state. This reduces, but not completely solves, the precision problem.

If a given $n$-qubit Hamiltonian contains multi-qubit interactions involving all of its constituents, that is, an $n$-qubit Hamiltonian with up to $n$-body terms, one would require $2^{\frac{n+2}{2}}-2$ total qubits to reduce all possible combinations of qubit interactions to two-body terms for an even $n$ ($3\times 2^{\frac{n-1}{2}}-2$ for odd $n$). This can be achieved by dividing the total qubit register into two fully-connected graphs using ancillary variables and connecting both graphs with another ancillary qubit.
Unfortunately, this will be the case for a general conversion from ANF to NNF due to the fact that an $n$ term sum in ANF will generally require 
\begin{equation}
    \sum_{k=1}^n\binom{n}{k}=2^n-1
\end{equation}
terms for the equivalent NNF equation.
Therefore one would need an exponential amount of quantum resources, ancillary qubits in this case, to encode the ground state into a Hamiltonian following this first direct approach.

\subsection{Truncated embedding}
\label{sec:truncated}

This problem can be circumvented by partitioning the original polynomials $p_i(\Vec{x})$ into smaller pieces with $k$-bounded length using ancillary variables. It is straightforward to see that a sum of $n_i$ monomials can be reduced to sums of up to $k$ terms by adding ancillas in the form
\begin{equation}
    \begin{split}
    x_1+\ldots+x_{n_i}=0 \rightarrow& x_1+\ldots +x_{k-1}+a_1=0 \\
    &a_1+x_k+\ldots +x_{2k-2}+a_2=0 \\
    &\ldots \\
    &a_l+x_{{n_i}-k+1}\ldots+x_{n_i} = 0\,,
    \end{split}
\end{equation}
at the cost of expanding the number of equations to $\frac{n_i-2}{k-2}$ using $l=\frac{n_i-2}{k-2}-1=\frac{n_i-k}{k-2}$ ancilla variables we have labeled $a_i$. A similar technique can also be used when encoding the $\mq$ problem to a SAT instance \cite{mq2sat}.

\begin{figure}[t]
\centering
\includegraphics[width=1.0\linewidth]{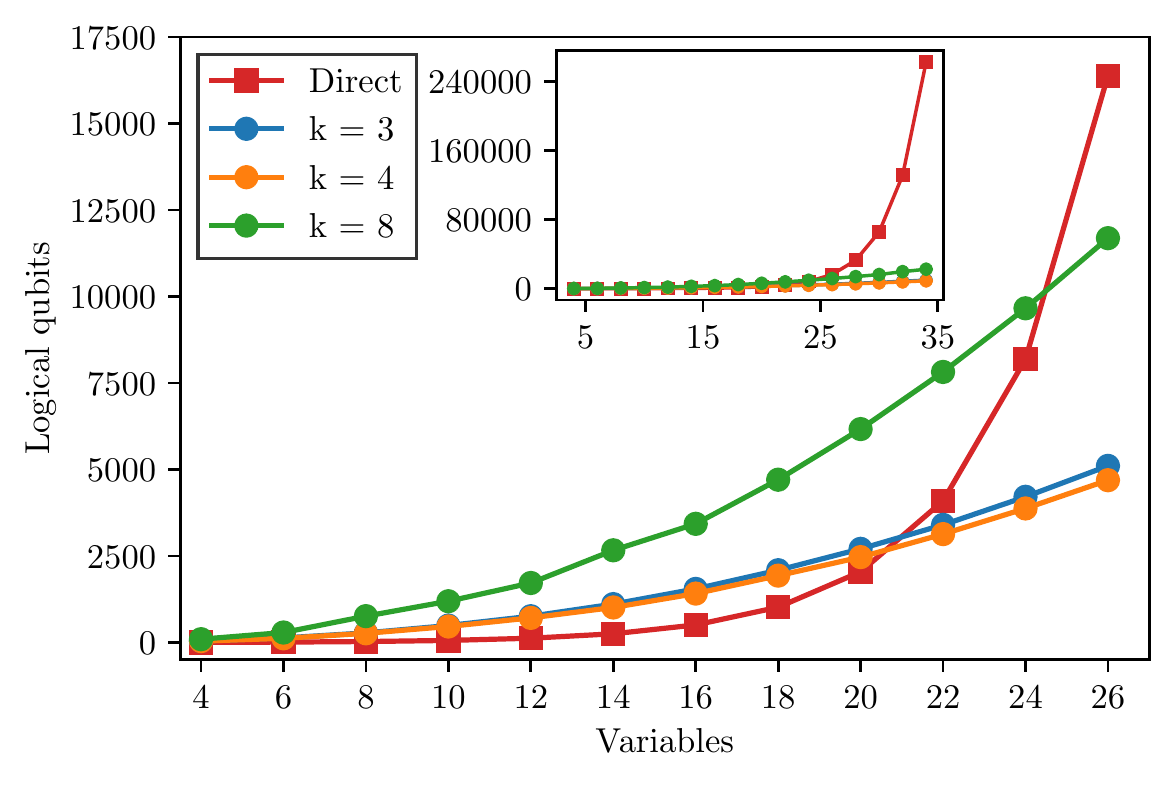}
\caption{Number of logical qubits needed to embed an $\mq$ problem into the ground state of a Hamiltonian for the direct and truncated approaches. It can be seen how the direct approach, while more resource-efficient in small systems, quickly outpaces the truncated approach. The inset shows the scaling for a larger number of variables. Note also that the optimal value for the cutoff variable is $k=4$.}
\label{fig:direct-vs-trunc}
\end{figure}

To have more precise control on the total number of ancillary qubits added to decompose the multi-qubit terms, we need to ensure that the parameter $k$ is also the maximum number of multi-body interactions. For that reason, we first introduce ancilla variables to substitute the two-qubit terms in the original ANF representation adding the required penalty functions. In the worst-case scenario, where all combinations of two-body interactions appear, we will need to add $\binom{n}{2}=n(n-1)/2$ ancillary variables. This, however, does not change the total number of monomials in the truncated system of equations.
The maximum number of qubits needed to represent an $\mq$ problem with $m$ equations involving $n$ variables into a two-body Hamiltonian will then be
\begin{equation}
    \sum_{i=1}^m \left[\frac{n_i-2}{k-2}\left(2^{\frac{k+2}{2}}-2-k\right)+\frac{n_i-k}{k-2}\right]+\binom{n}{2}+n,
\end{equation}
where $n_i$ is the number of monomials in each $p_i(\Vec{x})$ of Eq.~\ref{equ:system} and $k$ (even) is the length of the partitions. 

\begin{table*}[t!]
\centering\renewcommand\cellalign{c}
\setcellgapes{3pt}\makegapedcells
\begin{tabular}{|c|c|c|} \hline
\textbf{Gate} & \,\,\,\textbf{Boolean operation}\,\,\, & \textbf{Penalty function} \\ \hline
NOT & $z = \overline{x}$ & $2xz - x - z + 1$ \\ \hline
Controlled-NOT & $z = x_c x_t$ & $\makecell{2x_c x_t - 2(x_c + x_t)z - 4(x_c + x_t)x_a + 4zx_a + x_c + x_t + z + 4x_a}$\\ \hline
Toffoli & $z = x_{c1} x_{c2} x_t$ & $\makecell{-4x_{a1}x_{a2} + 4x_{a1}z - 4x_{a1}x_t - 2x_{a1}x_{c1} -\\ 2x_{a2}x_{c2} 2x_{a2}z + 2x_{a2}x_t + x_{c1}x_{c2} - 2x_t z + 4x_{a1} + 4x_{a2} + z + x_t}$\\ 
\hline\end{tabular}
\caption{Summary of Boolean operations and their penalty function implementation in a quantum annealer as only constant, single and two-qubit monomials appear in $\mq$ problems. The result is saved in the qubit corresponding to the variable $z$, while the variable $x$ corresponds to other qubits involved in the Boolean operation. The subscripts $c$, $t$, and $a$ correspond to \textit{control}, \textit{target}, and \textit{ancilla}, respectively.}
  \label{table:operations}
\end{table*}

The differences in scaling between the truncated approach with different values for $k$ and the direct embedding can be seen in Fig. \ref{fig:direct-vs-trunc}. While the direct approach is more efficient when the number of variables is small, its exponential scaling quickly makes it unfeasible when compared to the truncated approach. For the polynomial approach, the cutoff variable $k$ defines its scaling. We note that the scaling is optimal for $k=4$. 
Additionally, the precision issues raised in the substitution scheme will be significantly attenuated in the truncated embedding since now the non-locality of the ancillae is governed by $k$ and not the total number of variables.

A partition length of $k=4$ minimizes the total number of ancillae since the exponential term $2^{\frac{k+2}{2}}$ dominates and the truncation of the original equation needs half the parameters than $k=3$. Precisely, a $4$-term sum will only need $2$ extra ancillary variables to reduce it to up to $2$-body terms. Fixing the value for $k$ the total number of qubits needed reads
\begin{equation}
    \frac{n^2}{2}+\frac{n}{2}-4m+\frac{3}{2}\sum_{i=1}^mn_i.
\end{equation}
We encounter now a polynomial scaling with the number of parameters under the condition that the total number of terms in the system of equation scales reasonably with the number of variables.
In order to obtain a qubit scaling that only depends on the number of variables $n$, we can use average values for both $m$ and $n_i$. Generally, we will encounter as many equation as variables in the system, $m=n$, each with an average number of monomials given by the total possible combination of terms with two-body interactions, $n_i\sim \left(n+\binom{n}{2}\right)/2=(n+n^2)/4$. These two approximations yield the new scaling
\begin{equation}
    \frac{3}{8}n^3+\frac{7}{8}n^2-\frac{7}{8}n,
\end{equation}
a polynomial of degree $3$ in the number of variables of the problem.

\subsection{Penalty embedding}
\label{sec:problem_penalty}

An alternative way to embed the ground state of the $\mq$ problem is to model the equations in their ANF using logical quantum gates such as CNOT or Toffoli gates which natively act over the $\FF_2$ field, and then reproduce that circuit as an adiabatic evolution using penalty functions. To be more precise, we model the $\mq$ problem equations as Boolean operations on an output quantum register, that is, the actions of $+x_i$ and $+x_ix_j$ can be modeled to a CNOT and Toffoli gates targeting the output qubit and controlled by qubits $\{x_i\}$ and $\{x_i,\,x_j\}$ respectively. Then a Hamiltonian is constructed with a ground state that follows the correct gate-by-gate implementation of the resulting circuit.
This method of circuit-to-Hamiltonian encoding using penalty functions is reminiscent of Feynman's Hamiltonian clock \cite{feynman1985quantum}, where in order to create a Hamiltonian that faithfully represents the actions of a logical quantum circuit one would use an extra \textit{clock register} where the time step of each applied quantum gate is stored. In this implementation, an ancillary \textit{output} qubit register is added, which stores the result of the output qubit after each gate application.
The penalty functions needed to map the solution of an $\mq$ problem into the ground state of a Hamiltonian are displayed in Tab. \ref{table:operations}. The \textit{output} ancilla qubit $z$ used in the penalty function of a given quantum gate will be used as the \textit{target} qubit $x_t$ in the penalty function of the immediately following gate. These penalty functions contribute with positive energy if the state of the qubits involved does not match the logical Boolean operation that they map. Additionally, the qubits used to initialize the \textit{output} ancilla register are penalized if they are in the $\ket{1}$ state as we assume an initial state of the output qubit of $\ket{0}$. The same thing is applied to the \textit{output} ancillae where the final result of applying each equation $p_i(\Vec{x})$ in Eq.~\ref{equ:system} is stored as we are interested in the solution where the output is zero.
It is straightforward to see that the quantum resources needed to apply this implementation are governed by the number of monomials present in the equations of a given $\mq$ problem, as they will dictate the number of gates that are to be implemented. As discussed in Sec. \ref{sec:truncated} above, the average number of monomials appearing in a given problem will scale as $\order{n^3}$, and the ancilla overhead needed for the implementation of each CNOT or Toffoli gate, an extra $1$ or $2$ ancillae respectively, will not change the overall scaling. Therefore, up to the particularities of each implementation, both the truncated and the penalty function embedding will scale similarly, and in large system sizes outclass the direct embedding.
%


However, it is crucial to mention the number of physical qubits needed for the implementation when assessing the actual quantum resources. Due to chip architecture constraints, mapping a Hamiltonian into a real quantum annealing device will require an overhead to account for non-local interactions. The D-Wave API provides the automatic solver \textit{minorminer}~\cite{minorminer} to find a good embedding into their architecture. Highly non-local Hamiltonians will require a large number of physical qubits in order to represent each logical variable. Moreover, the amount of required physical qubits can change the scaling of a particular method, giving an edge to a more local embedding with more logical variables.

We show in Tab.~\ref{table:log-phys} the comparison between both the truncated and penalty embeddings in terms of physical and logical qubits required for their implementation into the Advantage D-Wave machine. For different instances up to 12 logical variables, we show the amount of required physical quantum resources for both the truncated and the penalty embedding. Each embedding has been averaged over $10$ instances in order to reduce the uncertainty due to the minimization method provided by D-Wave.
We note that both the truncated and the penalty embedding scale in a similar manner when mapped into the physical qubits of a given chip architecture with a worse, albeit still polynomial, overall scaling.
\begin{table}[t!]
\centering\renewcommand\cellalign{c}
\setcellgapes{3pt}\makegapedcells
\begin{tabular}{|c||c|c|c|c|c|} \hline
Variables & 4 & 6 & 8 & 10 & 12 \\ \hline \hline
Truncated (logical) & $30$ & $90$ & $231$ & $451$ & $718$ \\ \hline
Truncated (physical) & $55.6$ & $223.0$ & $758.0$ & $1627.8$ & $2645.2$ \\ \hline \hline
Penalty (logical) & $61$ & $150$ & $345$ & $645$ & $1005$ \\ \hline
Penalty (physical) & $105.1$ & $309.4$ & $864.1$ & $1940.6$ & $3436.5$ \\
\hline\end{tabular}
\caption{Number of logical and physical qubits needed to map the truncated and penalty embedding for different number of variables. Physical qubit values are averaged over 10 instances of the \textit{minorminer} algorithm provided by the D-Wave API \cite{minorminer}.}
\label{table:log-phys}
\end{table}

\section{Results}
\label{sec:experiments}

In this section, we encode some reduced $\mq$ problem instances using the methods presented in Sec. \ref{sec:formalizing} in order to be solved using D-Wave machines. We also propose an iterative method to aid in finding a singular correct solution in large, highly correlated, systems. For a detailed implementation, we refer to our code made available on Github \cite{nonlinear-code}. 


So far we have presented several ways to encode the solution of an $\mq$ problem into the ground state of a Hamiltonian that can be used for quantum annealing. The implementation of such a protocol on a real quantum device, however, will require adjusting to the specifics of the particular machine. The problem Hamiltonian assumes the implementation of all-to-all interaction. However, this is unrealistic because the superconducting chips for quantum annealing provided by D-Wave's Advantage device support a Pegasus chip architecture~\cite{dattani2019pegasus} and, therefore, the qubits need to be mapped accordingly to that restriction. The solution to this problem is the introduction of \textit{qubit chains}.
A logical qubit will be extended into a chain of qubits when mapped into the physical chip of the quantum device. This means that different physical qubits, which will represent the same variable, are bound together by an interaction term, a \textit{chain strength}, that penalizes members of the same qubit chain for being in different quantum states. We leave the mapping of the original variables to physical objects to the built-in compiler provided by the D-Wave library \cite{minorminer} and adjust the chain strength hyper-parameter in order to not overpower the variables of the problem while measuring as few broken chains as possible. This mapping will result in a more complex evolution, and consequently poorer results, especially for Hamiltonians with a large number of non-local qubit interactions.

\begin{figure*}[t]
    \centering
    \includegraphics[width=1.0\linewidth]{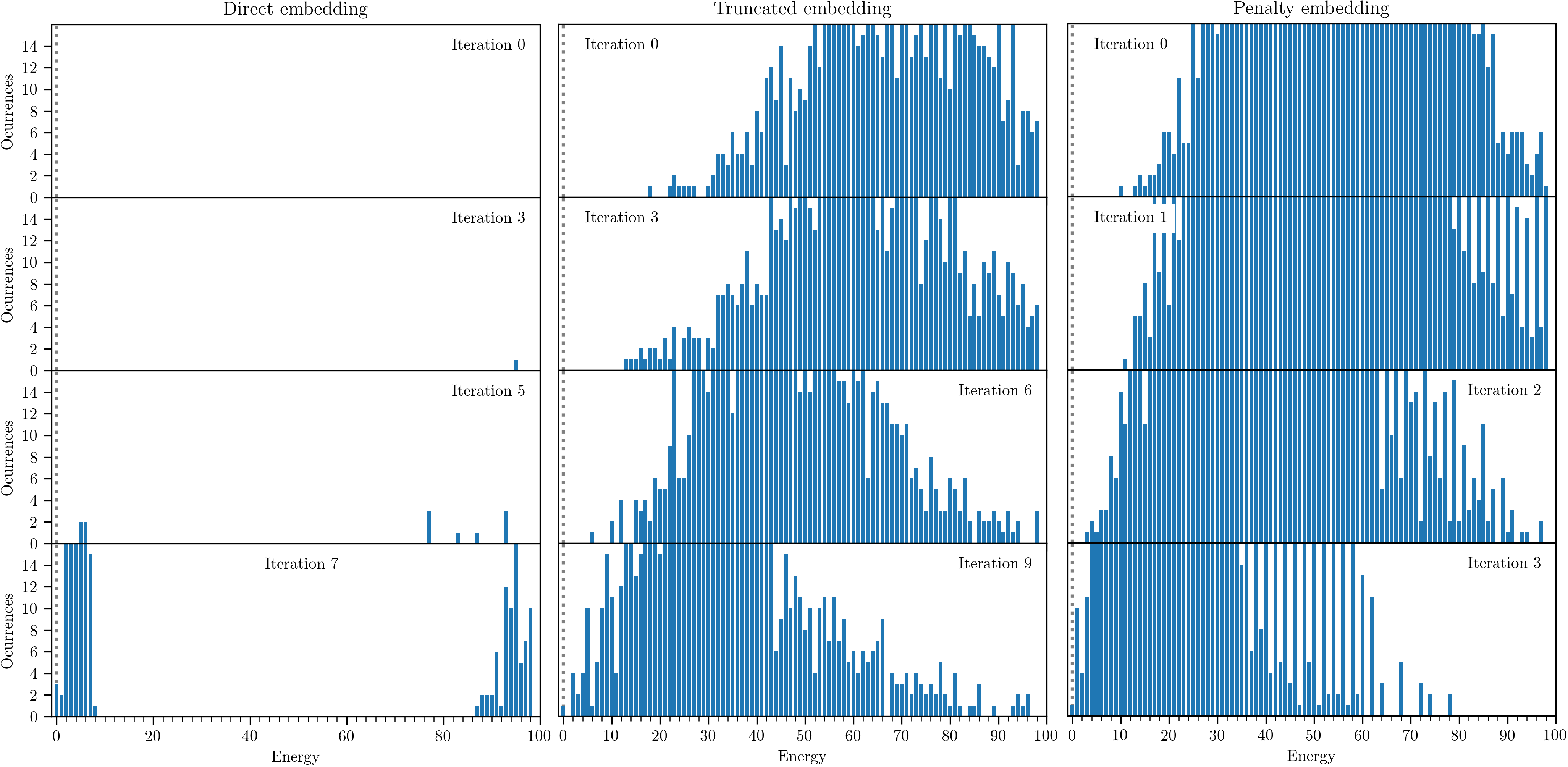}
    \caption{Samples below 100 energy after a 1000-sample execution on D-Wave's Advantage machine for the different $\mq$ problem embeddings presented where the ground state was first observed. We have implemented problems with a similar number of physical qubits required in the first embedding of the problem. The direct embedding (left) encodes a 9-variable problem in 46 logical qubits that are mapped to 179 physical ones, the truncated (center) and penalty (right) embedding both encode a 5-variable problem with 67 logical qubits with 167 physical ones for the truncated and 114 logical qubits with 221 physical one for the penalty one. We show 4 different iterations in our iterative method to highlight how the energy approaches the ground state as the system gets smaller both in logical and physical qubits. The ground state energy of this problem is 0, depicted as a gray dotted line.}
    \label{fig:embedding_results}
\end{figure*}

We present the results of running the Hamiltonians proposed in the different embedding schemes. The uppermost graphs in Fig. \ref{fig:embedding_results} show the results of sampling the final state of a quantum annealing evolved under each corresponding Hamiltonian for the direct, penalty, and truncated embedding, respectively. We state for each case the number of logical and physical qubits the problem needed to be mapped to. We decided to focus on a similar number of physical qubits needed, therefore the direct embedding was able to reach a 9-variable problem while the truncated and penalty embedding are limited to 5 variables. We note that with a small number of variables, the annealing process does not yield the exact ground state that encodes the solution of the problem, the quantum state with zero energy. Longer annealing times, more precise control on the annealing schedule, or higher quality qubits are ways to improve the results. However, current quantum annealers might not have the capabilities of tuning those parameters to the required specifications of large problems. In order to achieve the ground state energy of the problem in a machine-agnostic way, we propose an iterative algorithm that closes in on a smaller, easier-to-solve, subspace where the ground state might be located.

As detailed in Sec. \ref{sec:formalizing}, the proliferation of ancillary qubits in the different proposed embedding options appears when reducing the multi-qubit interactions into at most two-qubit interaction terms. This means that most of these added ancillae will represent products of other variables and will therefore have a stronger penalization than the original variables that they represent. That is, the wrong state of certain ancillary variables is penalized with a higher amount of energy than others. The following is a heuristic iterative method where we use that to our advantage.

After running an annealing protocol on a quantum device, if no quantum state with zero energy has been found, we may look at some of the low energy configurations of the obtained samples. If some qubits are found in the same result in all of the lowest energy states, we can assume that the Hamiltonian penalizes those variables more than the others. We can narrow the subsequent search space by substituting that variable in the original Hamiltonian by their, now known, preferred value. The more the search space is reduced, the easier it is for the quantum annealing device to find the lowest energy solution.

The amount of low-energy solutions to check for the same value of the ancilla variables is a hyper-parameter that can be optimized. On the one hand, if we set the value too low we might be excluding the ground state from the reduced search space by fixing ancillae to a wrong outcome. On the other hand, if we set it too high we might not find any variable that lays in the same output for all low energy configurations. More sampling at each iteration will also enhance our ability to fix ancillas but will impact the overall run time of the algorithm. The number of fixed parameters per run will depend on the problem, encoding and quality of the quantum device. We used a heuristic approach when tuning the number of low energy solutions checked. A method to check if the reduced subspace no longer contains the solution can be devised. If the lowest energy sample of the reduced Hamiltonian is lower than its equivalent value in the original Hamiltonian, adding back up the fixed variables of the original setting, then we have excluded the original ground state from the reduced subspace. The different rows in Fig. \ref{fig:embedding_results} show how this iterative approach indeed helps in finding the ground state of the problem. As more iterations go by and more ancillae are fixed, the system starts finding lower energy solutions until the ground state with zero energy is reached. 

The first row in Fig. \ref{fig:embedding_results} is always the initial run of the algorithm. Then the following are some snapshots of the energy samples during the iterative algorithm, and the last row showcases the first iteration where a state with energy zero is reached. The direct embedding shows the result of having a highly non-linear Hamiltonian with large parameters. The initial runs are very far away from the ground state and it is not until the more volatile variables are fixed that the ground state can be found. The truncated and penalty embedding behave in a similar way to each other. It can be seen how after each iteration the median energy gets closer and closer to the ground energy until it is reached. We note that the penalty embedding in spite of requiring more qubits to embed the problem Hamiltonian, reaches the ground state with fewer iterations. This can be attributed to the lower coefficients that are needed to map the problem, making it more suited to an annealer machine such as the one provided by D-Wave.

\section{Conclusion}
\label{sec:conclusions}

Our work is a first step towards demonstrating the efficiency of quantum annealing computations in solving the $\mq$ problem, using practical experiments on the existing D-Wave quantum annealing platform. 
We show that we can construct a Hamiltonian with a ground state encoding the solution of the problem and subsequently find it using a quantum annealer. We propose different methods for the embedding of the problem into a Hamiltonian using a polynomial amount of quantum resources.
As quantum technology advances, we foresee that the evolution of quantum annealing architectures (e.g., support to n-body interactions or larger coherence times) might provide a quantum advantage when solving such problems as the required numbers of ancilla qubits and required quantum control would decrease.

We have introduced an algorithm that simplifies the problems by fixing ancillary qubits that are easy to find for the quantum device in order to more reliably find the ground state of the more complex qubits with finer parameters. This method can help when dealing with large amounts of qubits in near-term devices and could be applied in problems beyond the scope of what is studied in this manuscript.

As an estimate of the quantum resources needed to solve state-of-the-art $\mq$ problems, we refer to the Fukuoka MQ Challenge website \cite{mqchallenge} where the largest unsolved instances of $\mq$ problems can be found. The Type I challenge problem instance of 37 equations and 74 variables would require an estimate of under 80000 logical qubits for its solution to be mapped into the ground state of a two-body Hamiltonian. For the Type IV challenge with 69 equations and 105 variables, one would require under 300000 logical qubits for the embedding. Quantum annealers are still far away from being able to tackle the problems at the edge of what is classically solvable, but quantum technologies are still emerging and new devices with more, and higher quality, qubits are being currently developed.


\section*{Acknowledgments}
The authors would like to thank Andre Esser for his insights. M.M. is partially supported by the TRUSTIND project, under the grant agreement KK-2020/00054, from the Department of Economic Development and Infrastructures of the Basque Government. M.M. is also a member of the Intelligent Systems for Industrial Systems research group of Mondragon Unibertsitatea (IT1676-22), supported by the Department of Education, Universities and Research of the Basque Country.


\bibliography{Citations}

\end{document}